# Initiation of nuclear reactions under laser irradiation of Au nanoparticles in the aqueous solution of Uranium salt

## A.V. Simakin and G.A. Shafeev

Wave Research Center of A.M. Prokhorov General Physics Institute of the Russian Academy of Sciences, 38, Vavilov street, 119991 Moscow Russian Federation

## Abstract

Laser exposure of suspension of either gold or palladium nanoparticles in aqueous solutions of $UO_2Cl_2$ of natural isotope abundance was experimentally studied. Picosecond Nd:YAG lasers at peak power of $10^{11}$ -$10^{13}$ W/cm$^2$ at the wavelength of $1.06 - 0.355$ μm were used as well as a visible-range Cu vapor laser at peak power of $10^{10}$ W/cm$^2$. The composition of colloidal solutions before and after laser exposure was analyzed using atomic absorption and gamma spectroscopy in $0.06 - 1$ MeV range of photon energy. A real-time gamma-spectroscopy was used to characterize the kinetics of nuclear reactions during laser exposure. It was found that laser exposure initiated nuclear reactions involving both $^{238}$U and $^{235}$U nuclei via different channels in $H_2O$ and $D_2O$. The influence of saturation of both the liquid and nanoparticles by gaseous $H_2$ and $D_2$ on the kinetics of nuclear transformations was found. Possible mechanisms of observed processes are discussed.

PACS numbers: 42.62.–b; 61.46.+w; 78.66.–w

*Corresponding author, e-mail shafeev@kapella.gpi.ru

## Introduction

Modern lasers allow excitation of nuclear energy levels via generation of high-energy particles that appear during the interaction of laser radiation with plasma produced on a solid target. Successful excitation of nuclear levels has been reported for some isotopes of Hg and Ta under exposure of a target in vacuum to a femtosecond laser radiation. [1,2]. Emission of gamma-photons from a Ta target exposed in vacuum to peak intensity of $10^{18}$ W/cm$^2$ in a femtosecond range of pulse duration results has been reported recently [3]. The average energy of these photons is about few MeV. Picosecond laser plasma is also a source of high-energy particles whose energy is sufficient for excitation of energy levels of nuclei in the exposed target [4].

Another possibility for laser excitation of nuclear energy levels consists in laser exposure of nanoparticles suspended in a liquid (colloidal solution). This scheme allows laser initiation of nuclear reactions, e.g., transmutation of $^{196}$Hg into $^{197}$Au [5, 6] via laser exposure of Hg nano-drops in heavy water $D_2O$. It is believed that thermal neutrons needed for this transmutation are released from Deuterium though the mechanism of this release remained unknown. The possibility to induce nuclear reactions at relatively low peak intensity of laser radiation was attributed to the local field enhancement in the vicinity of metallic nanoparticles by a factor of $10^4$-$10^5$. This may provide effective peak intensity in the liquid of about $10^{17}$ W/cm$^2$, which is already comparable with those used for exposure of solid targets in vacuum.

It is of interest to use the same approach for initiation of nuclear reactions in nanoparticles of unstable elements, such as $^{238}$U or $^{232}$Th. However, these elements are chemically reactive and would react with aqueous environment during the laser synthesis. The solution of this problem consists in using NPs of noble metals, e.g., Au, to provide the constant level of absorption in the liquid, while unstable elements can be presented in the solution as aqua-ions [7].

The aim of this work is the experimental study of possibility of laser initiation of nuclear reaction in aqueous solutions of a Uranium salt under absorption of laser radiation by Au nanoparticles. The most known Uranium isotopes are $^{238}$U and $^{235}$U. They undergo the sequence of α- and β-decays as follows:

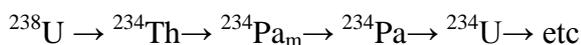

$$^{238}\text{U} \rightarrow {}^{234}\text{Th} \rightarrow {}^{234}\text{Pa}_m \rightarrow {}^{234}\text{Pa} \rightarrow {}^{234}\text{U} \rightarrow \text{etc}$$

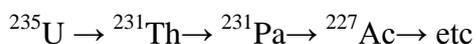

$$^{235}\text{U} \rightarrow {}^{231}\text{Th} \rightarrow {}^{231}\text{Pa} \rightarrow {}^{227}\text{Ac} \rightarrow \text{etc}$$

Raw Uranium contains 0.7% of another isotope $^{235}$U that originates from $^{235}$Np. Natural decay time of both Uranium isotopes is very long ($4.5\times10^9$ and $7\times10^8$ years, respectively). One may expect that thermal neutrons generated through laser exposure of Au NPs in aqueous solutions should alter the equilibrium concentration of all elements that belong to U branching.

**Experimental**

Au nanoparticles (NPs) were synthesized by ablation of a bulk gold target either in $H_2O$ or $D_2O$ with the help of a Nd:YAG laser with pulse duration of 70 ns at wavelength of 1.06 μm. The details of the synthesis can be found elsewhere [8, 9]. The resulting average size of Au NPs as determined by Transmission Electron Microscopy lies between 10 and 20 nm. The Uranium salt $UO_2Cl_2$ of natural isotope composition was then dissolved in the colloidal solution, and the solution was divided into two parts, one of them considered as the initial solution. The second part of the solution was exposed to laser radiation. The exposure was carried out either of the as-obtained solution or under continuous purge of $H_2$ or $D_2$ for $H_2O$ and $D_2O$, respectively. The gases were obtained by electrolysis of corresponding liquids, either $H_2O$ or $D_2O$ and were supplied to the solution at atmospheric pressure.

Three laser sources were used for exposure on Au NPs in the aqueous solutions of the Uranium salt. These were a Nd:YAG laser, pulse duration of 150 ps, wavelength of either 1.06 or 0.355 μm, energy per pulse of 100 at 1.06 and 20 mJ at 0.355 μm, repetition rate of 10 Hz, peak power of $10^{13}$ W/cm$^2$, a Nd:YAG laser, pulse duration of 350 ps, wavelength of 1.06 μm, energy/pulse of 350 μJ, repetition rate of 300 Hz, peak power of $10^{11}$ W/cm$^2$, and a Cu vapor laser, pulse duration of 10 ns, wavelength of 510/578 nm, energy/pulse of 100 μJ, repetition rate of 15 kHz, peak power of $10^{10}$ W/cm$^2$.

Gamma-emission from samples before and after laser exposure was characterized using a semiconductor γ-spectrometer Ortec-65195-P. This provided the analysis of sample specific activity in γ-photons from 0.06 до 1.5 MeV in Becquerel per ml. Real-time acquisition of γ-spectra of the solutions during laser exposure was achieved with the help of a portable scintillator γ-spectrometer. In the latter case the cell with the solution was fixed just on the spectrometer itself, which guaranteed the constant geometry of measurements under natural background of γ-radiation. The acquisition time was sufficiently long to provide the accuracy of measurements better than 3% in the channel with maximal number of counts indicated by the spectrometer.

## Results and discussion

Exposure of Au NPs in aqueous solutions of $UO_2Cl_2$ either in $H_2O$ or in $D_2O$ leads to significant modifications of the activity of all elements of U branching. The result of the laser exposure depends on the kind of water used in the experiment. Exposure in $D_2O$ results in the decrease of the activity of both Uranium isotopes at laser peak power of $10^{10}$-$10^{11}$ W/cm². Activity is linearly related to the quantity of the corresponding isotopes therefore, one may conclude that laser exposure of Au NPs in presence of aqua-ions of $UO_2^{-2}$ leads to the accelerated decay of $^{238}$U.

In case of laser exposure of Au NPs in $H_2O$ with $UO_2Cl_2$ the result is the opposite. $^{238}$U is not gamma-active, and the modifications of its concentration can be inferred from the activity of its daughter nuclides, $^{234}$Th and $^{234}$Pa. The activity of $^{234}$Th and $^{234}$Pa$_m$, as well as $^{234}$Pa increases after laser exposure (see Fig. 1, a). Note that these elements are daughter ones for $^{238}$U. In Fig. 1, a  one can see that the activity (concentration) of $^{231}$Th also increases after laser exposure. The parent of this element is $^{235}$U, and the increase of $^{231}$Th signifies its accelerated decomposition. However, the concentration of $^{235}$U increases after laser exposure as it is shown in Fig. 1, b.

It is pertinent to note that no measurable changes of the activity of nuclides of U branching were detected under exposure of the colloidal solutions of Au NPs in either $H_2O$ or $D_2O$ with $UO_2Cl_2$ with radiation of a femtosecond radiation of a Ti:sapphire laser at peak power of $10^{13}$ W/cm² at wavelength of 800 nm.

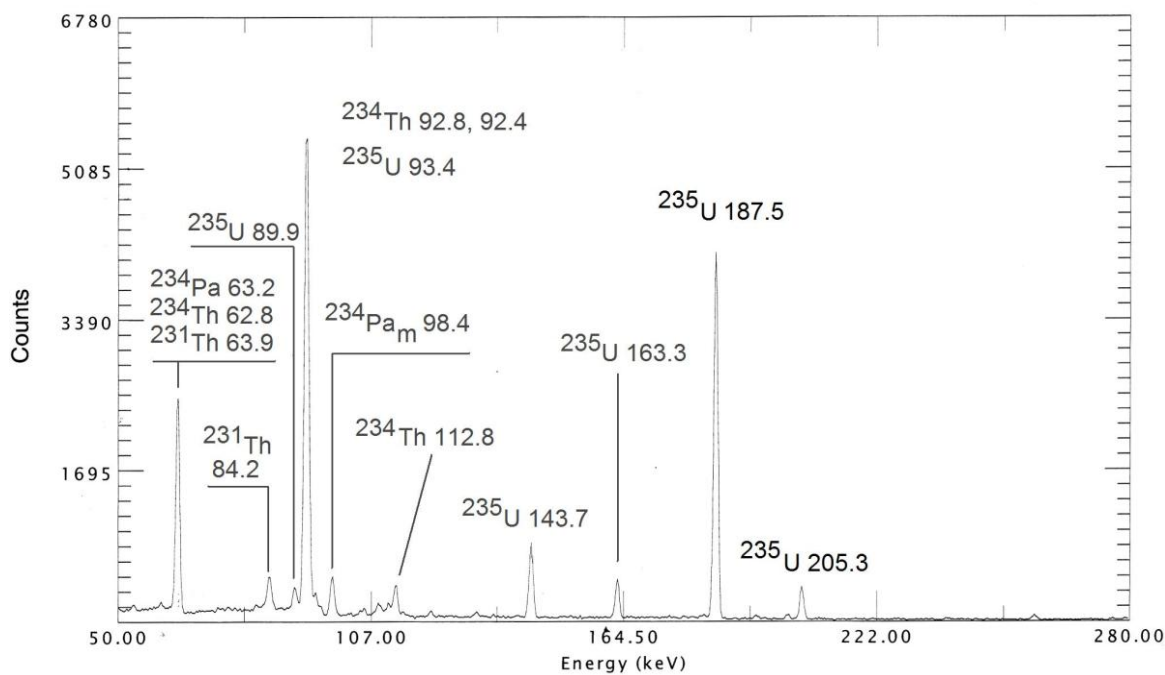

a

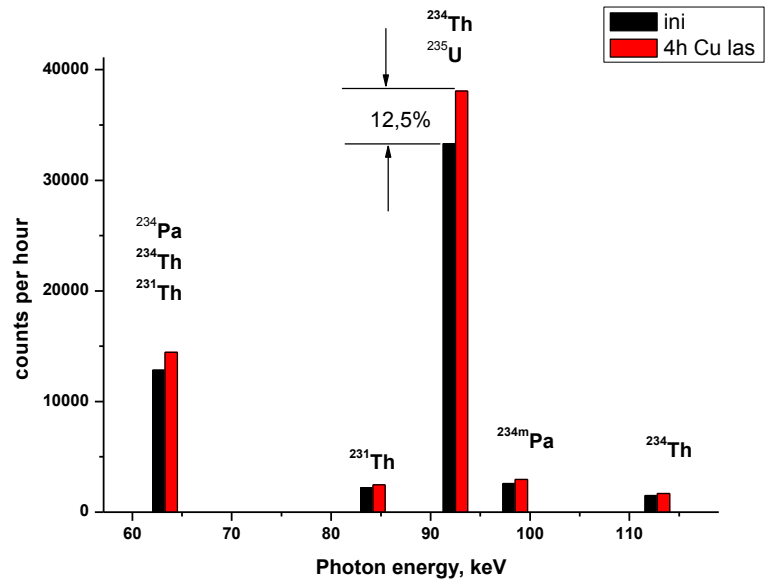

b

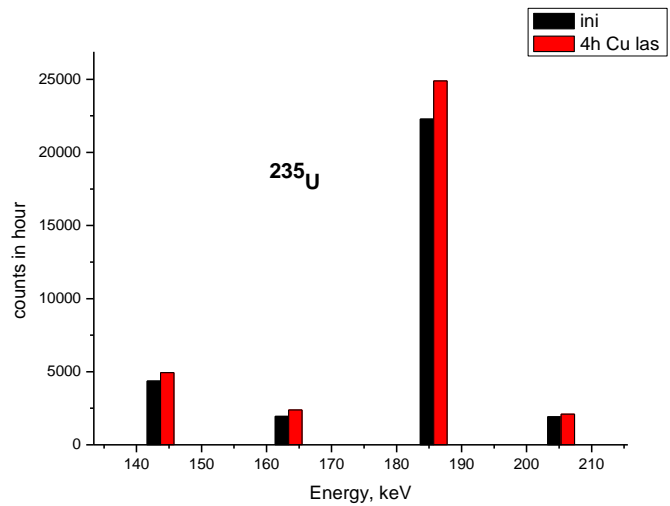

c

Fig. 1. Gamma-spectrum of the initial solution of $UO_2Cl_2$ in $H_2O$ with Au NPs (a). Gamma-spectrum of elements of $^{238}U$ branching before (black) and after laser exposure (red) of the colloidal solution of Au NPs in $H_2O$ with $UO_2Cl_2$ (b). Gamma-spectrum of $^{235}U$ before (black) and after (red) laser exposure of the colloidal solution of Au NPs in $H_2O$ with $UO_2Cl_2$ (c). Cu vapor laser 4 hours of exposure, peak power of $10^{10}$ W/cm$^2$, repetition rate of 15 kHz.

Real-time γ-spectra of the samples are presented in the Fig. 2.

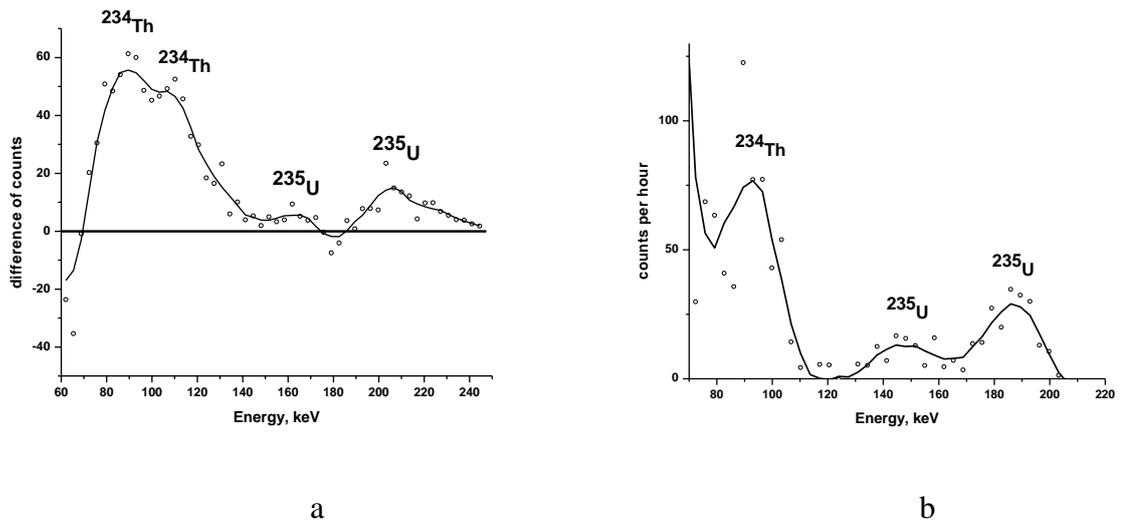

a                      b

Fig. 2. Differential spectra of the samples of Au NPs exposed to 350 ps laser radiation in $H_2O$ with purged $H_2$ (a) and in $D_2O$ with purged $D_2$ (b). Initial spectra of the same sample are subtracted in each case.

One can see that the samples are characterized by γ-emission of the nuclides belonging to $^{238}U$ branching as well as by that of $^{235}U$. However, different peaks of these nuclides are active under laser exposure in $H_2O$ and $D_2O$ at otherwise equal conditions.

The tendency changes at higher peak power of the laser radiation. Namely, at the peak power of order of $10^{13}$ W/cm$^2$ in 150 ps pulses the activity of both U isotopes increases after laser exposure of the colloidal solution of Au NPs in $D_2O$. This is illustrated in Fig.3.

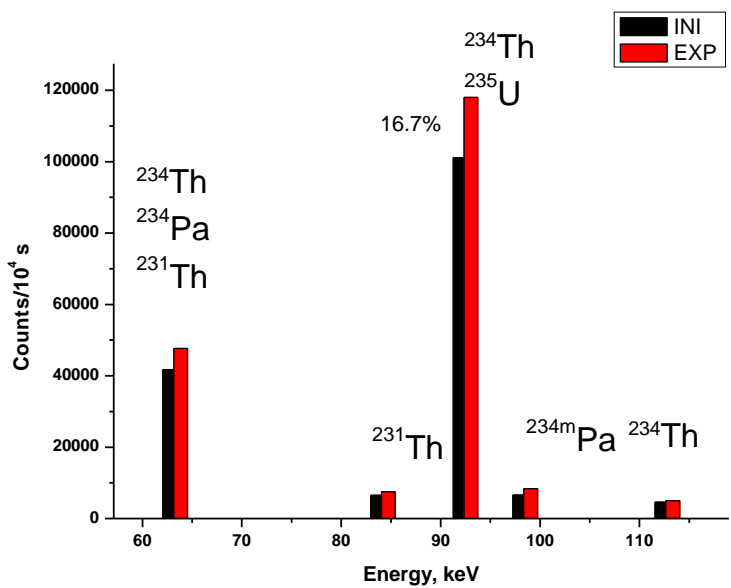

Fig. 3. Gamma-spectrum of the sample of $UO_2Cl_2$ in $D_2O$ exposed to the first harmonics of a Nd:YAG laser, pulse duration of 150 ps, 1 hour of exposure at 10 Hz.

The activity of both U isotopes increases after the laser exposure of Au NPs in $D_2O$, as it can be concluded from the increase of activity of corresponding daughter nuclides.

The kinetics of the nuclear transformations is also sensitive to the laser wavelength. The dependence of the activity of several nuclides of U branching on the concentration of $UO_2Cl_2$ is presented in the Fig. 4.

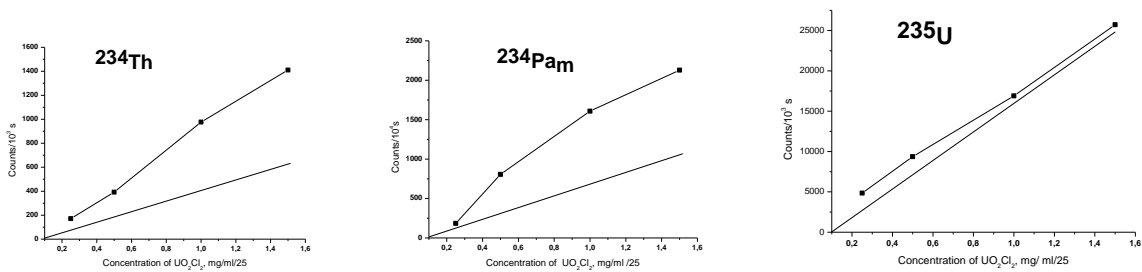

Fig. 4. Dependence of activity of $^{234}$Th, $^{234}$Pa$_m$, and $^{235}$U of the same probes exposed to the 3$^{rd}$ harmonics (0.355 μm) of a 150 ps Nd:YAG laser in $H_2O$ for 1 hour at 10 Hz repetition rate on the concentration of Uranium salt. Straight lines represent the activity of the same nuclides in the initial solution.

In this case the laser action is characterized by high selectivity. Indeed, the decay of $^{238}$U is noticeably accelerated by laser exposure of Au NPs along the branch $^{238}$U → $^{234}$Th→ $^{234}$Pa$_m$→ $^{234}$Pa, and the activity of $^{234}$Th in the laser-exposed sample is twice higher than in the initial sample. On the contrary, the activity (and related to it concentration) of $^{235}$U remains almost constant in the same probes.

Different reaction pathways observed under exposure in $H_2O$ and $D_2O$ imply different interaction of these compounds with Au NPs. This interaction is not related to chemical one since chemical properties of these two waters are the same. Indeed, NPs are molten during their synthesis by laser ablation and ionized during laser exposure. The emission of atomic Au has been detected under exposure of Au NPs in water at laser peak power of $10^{11}$ W/cm$^2$ at 1.06 mm wavelength. The upper electronic level of this emission is 5 eV, which is comparable with the energy of dissociation of water molecules (13.6 eV) [9]. Accordingly, the water vapor around the NPs is partially dissociated. Molecular gases $H_2/D_2$ dissolve in the metal while the solubility of O is much lower than that of H/D due to larger size. This process is very efficient in view of high specific surface of Au NPs used in this work since their surface is as high as 10 m$^2$ per 1 ml of colloidal solution. Saturation of the liquid with $H_2/D_2$ increases the quantity of these gases in Au NPs. If the solidification rate of NPs is sufficiently high, then the dissolved gases remain inside the NPs. Saturation of the liquid with $H_2/D_2$ increases the quantity of these gases in Au NPs. If the solidification rate of NPs is sufficiently high, then the dissolved gases remain inside the NPs.

Each nanoparticle can be considered as a target that is ionized by the laser pulse. The expansion of the plasma around the nanoparticle is confined by surrounding liquid, so that sufficiently long laser pulse can still interact with these nano-sized plasma entities.

## Conclusion

Further interpretation of the observed results on laser initiation of nuclear reactions cannot be performed on the basis of known phenomena. It seems that the gases dissolved in Au NPs provide the particles that further induce the nuclear reactions. The mechanism of the formation of these particles, most probably neutrons, remains unknown so far. However, the present results allow the conclusion that the energy spectrum of these neutrons depends on the number of experimental parameters, such as the nature of the aqueous environment, laser wavelength, peak power of laser radiation, etc. The mechanism of the initiation of nuclear reactions at relatively weak laser intensities of $10^{13}$ W/cm$^2$ requires further multi-parametric studies.

## Acknowledgements

The work was partially supported by Russian Foundation for Basic Research, grants ## 07-02-00757, 08-07-91950, and by Scientific School 8108.2006.2. Dr. A.V. Goulynin is thanked for gamma-measurements and helpful discussions.